\documentstyle[epsfig]{aipproc}

\begin{document}
\title{A Coordinated Radio Afterglow Program}

\author{D. A. Frail$^1$, S. R. Kulkarni$^2$, M. H. Wieringa$^3$, 
G. B. Taylor$^1$, G. H. Moriarty-Schieven$^4$, 
D. S. Shepherd$^1$,  R. M. Wark$^3$, R. Subrahmanyan$^3$, 
D. McConnell$^3$, and S. J. Cunningham$^3$}

\address{$^1$National Radio Astronomy Observatory, Socorro, NM
  87801\thanks{The NRAO is a facility of the National Science
    Foundation, operated under cooperative agreement by Associated
    Universities, Inc.}\\
$^2$Division of Physics, Mathematics, and Astronomy 105-24,
               Caltech, Pasadena, CA 91125\\
$^3$Australia Telescope National Facility, CSIRO,
              Epping 2121, Australia\\
$^4$Joint Astronomy Centre, 600 A'Ohoku Place, Hilo, HI
  96720
}

\maketitle

\begin{abstract}
  We describe a ground-based effort to find and study afterglows at
  centimeter and millimeter wavelengths.  We have observed all
  well-localized gamma-ray bursts in the Northern and Southern sky
  since BeppoSAX first started providing rapid positions in early
  1997. Of the 23 GRBs for which X-ray afterglows have been detected,
  10 have optical afterglows and 9 have radio afterglows. A growing
  number of GRBs have both X-ray and radio afterglows but lack a
  corresponding optical afterglow.
\end{abstract}

\section*{Introduction}

BeppoSAX revolutionized gamma-ray burst (GRB) astronomy not only
through its discovery of X-ray afterglows but also through the
dissemination of accurate and timely GRB positions to ground-based
observers, who then conduct searches of afterglows at optical and
radio wavelengths. Our collaboration uses the interferometer
facilities of the Very Large Array (VLA), the Australia Telescope
Compact Array (ATCA), the Very Long Baseline Array (VLBA) and the
Owens Valley Radio Observatory (OVRO) Interferometer.  At high
frequencies, we use single dish telescopes which include the James
Clerk Maxwell Telescope (JCMT) and the OVRO 40-m Telescope. All
afterglow searches begin with the VLA in the northern hemisphere
(dec.$>-45^\circ$, $\sigma_{\rm rms}=45$ $\mu$Jy in 10 min.,
FOV$\simeq5^\prime$) and the ATCA in the southern hemisphere
(dec.$<-45^\circ$, $\sigma_{\rm rms}=45$ $\mu$Jy in 240 min.,
FOV$\simeq5^\prime$), typically at a frequency of 8.5 GHz, which
provides a balance between sensitivity and field-of-view.  Follow-up
programs at the other radio facilities are begun after a VLA or ATCA
transient is discovered. 

As with quasars, radio observations provide unique diagnostics
complementary to those obtained at X-ray and optical wavelengths.  Our
collaboration has discovered all known radio afterglows to date,
leading to a number of important results: the direct demonstration of
relativistic expansion of the ejecta (Frail et al. 1997a), evidence
for a reverse shock (Kulkarni et al. 1999), the first true calorimetry
of a GRB explosion (Frail, Waxman \& Kulkarni 1999), the discovery of
optically obscured events (Taylor et al. 1998), the first unambiguous
evidence that the ejecta are collimated in jets (Harrison et al.
1999), and the discovery of a possible link between supernovae and
GRBs (Kulkarni et al. 1998).

\begin{figure}
  \centerline{\epsfig{file=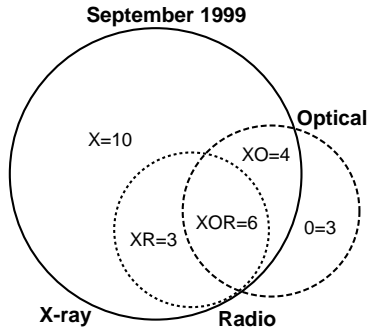,width=8.cm}}
  \vspace{-2.truein}
\caption{A Venn diagram showing the detection statistics for 26
  well-localized GRBs. Of the 23 GRBs for which X-ray afterglows have
  been detected to date, 10 have optical afterglows (XO + XOR) and 9
  have radio afterglows (XR + XOR). In total there are 13 optical
  and/or radio afterglows with corresponding X-ray afterglows (XO + XR
  + XOR). Only 6 GRBs have afterglows detected in all three bands
  (XOR).
\label{fig1}}
\end{figure}

\begin{table}
\caption{Radio Afterglow Detections.}
\label{table1}
\begin{tabular}{crrrll}
 \omit & {F$_{\rm peak}$ } & {t$_{\rm max}$} 
& {AG } & \omit & \omit \\
 {GRB} & {($\mu$Jy) } & {(days)} 
& {Class} & Instruments & References \\
\tableline
970508  &  1200 & 450     & XOR & VLA, VLBA, OVRO, JCMT &  Frail et al.~(1997a)  \\
970828  &  150  & 3.5     &  XR & VLA                   & Djorgovski et al.~(1999)  \\
980329  &  300  & 135     & XOR & VLA, OVRO, JCMT       & Taylor et al.~(1998)  \\
980425\tablenote{Related to SN1998bw. We do not include this GRB in the detection
  statistics.} &  50,000  & $>$300   & XOR & ATCA, JCMT        & Kulkarni et al.~(1998)   \\
980519  &  300  & 65      & XOR & VLA, OVRO             & Frail et al.~(1999a)  \\
980703  & 1200  & 210     & XOR & VLA, VLBA, JCMT       & Bloom et al.~(1998)  \\
981226  &  170  & 20      & XR  & VLA                   & Frail et al.~(1999b)  \\
990123  &  260  & 1.2     & XOR & VLA, OVRO, OVRO 40-m, JCMT & Kulkarni et al.~(1999)  \\
990506  &  550  & $<$16   & XR  & VLA                   & Taylor et al.~(1999)  \\
990510  &  225  & 20      & XOR & ATCA                  & Harrison et al.~(1999)   \

\end{tabular}
\end{table}

\section*{Radio Afterglow Statistics}

Since 1997 we have observed 19 GRBs with the VLA and detected a total
of eight radio afterglows (see Figure \ref{fig1}, Tables \ref{table1}
and \ref{table2}).  The peak fluxes (F$_{\rm peak}$) of the detections
range from 1200 $\mu$Jy to 150 $\mu$Jy.  This small range of F$_{\rm
  peak}$ values suggests that our ability to detect radio afterglows
is severely limited by the sensitivity of the telescope. The
``lifetime'' (i.e.  t$_{\rm max}$) of the radio afterglows is
signal-to-noise limited but it is clear, at least among bursts of
comparable brightness, that t$_{\rm max}$ varies substantially.  Of
special note are the three GRBs (970828, 981226, and 990506) which
have no optical counterparts (i.e.  XR class). These may represent an
important group of GRBs whose optical emission is extincted by dust.

There are 11 GRBs for which a VLA search of the error box failed to
detect a radio afterglow (see Table \ref{table2}). The peak fluxes
given in the table are conservative upper limits for a radio afterglow
on a timescale of 1 to 30 days and at frequencies between 1.4 and 8.5
GHz.  These non-detections vary in quality depending on the size of
the error circle but most observations had sufficient sensitivity to
detect radio afterglows with fluxes comparable to those listed in
Table \ref{table1}.

There have been two radio afterglow detections made at the ATCA (see
Table \ref{table1}). The possible relation of GRB 980425 to SN1998bw
makes it a rather unusual event, so we do not include it in the
detection statistics. The upper limits of the six ATCA non-detections
in Table \ref{table4} were not sufficient to have detected the weaker
radio afterglows in Table \ref{table1}.

\begin{table}
\caption{VLA Afterglow Non-Detections.}
\label{table2}
\begin{tabular}{crrl}
 {GRB} & {F$_{\rm peak}$ ($\mu$Jy) } 
& {AG Class} & References \\
\tableline
970111  &  $<$300     & X  & Frail et al.~(1997b) \\
970228  &  $<$50      & XO & Frail et al.~(1998)  \\
970616  &  $<$150     & X  & IAUC \#6691  \\
970815  &  $<$50      & X  & IAUC \#6723 \\
971214  &  $<$25      & XO & Ramaprakash et al.~(1998) \\
971227  &  $<$50      & X  & \omit  \\
980326  &  $<$150     & O  & \omit \\
980613  &  $<$50      & XO & \omit   \\
981220  &  $<$125     & X  & GCN \#269  \\
990520  &  $<$125     & X  & \omit  \\
990704  &  $<$125     & X  & \omit  \\
\end{tabular}
\end{table}

\begin{table}
\caption{ATCA Afterglow Non-Detections.}
\label{table4}
\begin{tabular}{crrl}
 {GRB} & {F$_{\rm peak}$ ($\mu$Jy) } & {AG Class} & References\\
\tableline
970402  &  $<$300      & X  & \omit   \\
980109  &  $<$550      & \tablenote{GRB 990109 and 990217 were 
seen only in gamma-rays. 
No afterglows were detected at any wavelength.} & \omit \\
990217  &  $<$175      & $^{\rm a}$ & GCN \#266\\
990627  &  $<$125      & X   & GCN \#357 \\
990705  &  $<$100      & XO  & GCN \#376 \\
990712  &  $<$100      & O   & \omit   \\
\end{tabular}
\end{table}

\section*{Summary}

  In summary, our coordinated program has been very successful in
  detecting radio afterglows from GRBs. In particular:

\begin{itemize}

\item{Six gamma-ray bursts are seen at X-ray, optical and radio
    wavelengths (GRB 970508, GRB 980329, GRB 980519, GRB 980703, GRB
    990123, GRB 990510) (see Figure \ref{fig1}).}

\item{Of the 23 X-ray afterglows, nine have been detected at radio
    wavelengths (XOR + XR) for a rate of 39\%. At the VLA the
    detection rate is 8/19 or 42\%. The small range in the observed
    peak flux densities suggests that our ability to detect radio
    afterglows is mainly limited by the sensitivity of the telescopes
    (VLA and ATCA).}

\item{Of the 23 X-ray afterglows, ten have been detected at optical
    wavelengths (XOR + XO) or 43\%. The detection rate of
    well-localized GRBs is comparable at optical and radio
    wavelengths.}

\item{There exists a growing class (XR) of ``dark'' GRBs which have
    X-ray and radio afterglows but no known optical afterglow. These
    may represent an important group of GRBs whose optical emission is
    extincted by dust.}

\end{itemize}

\end{document}